\documentclass[letterpaper, 10 pt, conference]{ieeeconf}  

\IEEEoverridecommandlockouts                              

\usepackage{cite}
\usepackage{amsmath,amssymb,amsfonts}
\usepackage{algorithmic}
\usepackage{graphicx}
\usepackage{textcomp}
\usepackage{xcolor}
\usepackage{url}
\usepackage{hyperref}
\usepackage{soul}
\usepackage{bm}
\usepackage{graphicx} 
\usepackage{booktabs} 
\usepackage{multirow} 
\usepackage{afterpage}
\usepackage{flushend}

\def\BibTeX{{\rm B\kern-.05em{\sc i\kern-.025em b}\kern-.08em
    T\kern-.1667em\lower.7ex\hbox{E}\kern-.125emX}}
\def\FMM{\text{Mammo-Mamba}}

\def\y{\bm{y}}
\def\Y{\bm{Y}}
\def\z{\bm{z}}
\def\y{\bm{y}}
\def\Z{\bm{Z}}
\def\h{\bm{h}}

\def\A{\bm{A}}
\def\B{\bm{B}}
\def\C{\bm{C}}

\def\x{\bm{x}}
\def\h{\bm{h}}
\def\X{\bm{X}}
\def\W{\bm{W}}
\def\G{\bm{G}}

\def\K{\bm{K}}

\def\P{\bm{P}}
\def\T{\bm{T}}
\def\b{\bm{b}}
\def\p{\bm{p}}

\def\ul{^{(\ell)}}

\def\i{_i}

\def\k{_{i, k}}
\def\pk{_{i, k-1}}
\def\ik{_{i,k}}
\def\mX{\bm{\mathcal{X}}}
\def\SMB{\text{SecMamba}}

\title{\LARGE \bf $\FMM$: A Hybrid State-Space and Transformer Architecture with Sequential Mixture of Experts for Multi-View Mammography}

\author{Farnoush Bayatmakou$^{1}$, Reza Taleei$^{2}$,
Nicole Simone$^{2}$,
Arash Mohammadi$^{1}$
\thanks{$^{1}$Concordia Institute for Information Systems Engineering (CIISE), Concordia University, Montreal, Canada. This work is partially supported by Natural Sciences and Engineering Research
Council (NSERC) of Canada through the NSERC Discovery Grant RGPIN-2023-05654.
}
\thanks{$^{2}$Thomas Jefferson University Hospital, Philadelphia, Pennsylvania, USA.}
}
\begin{document}

\maketitle
\begin{abstract}
Breast cancer (BC) remains one of the leading causes of cancer-related mortality among women, despite recent advances in Computer-Aided Diagnosis (CAD) systems. Accurate and efficient interpretation of multi-view mammograms is essential for early detection, driving a surge of interest in Artificial Intelligence (AI)-powered CAD models. While state-of-the-art multi-view mammogram classification models are largely based on Transformer architectures, their computational complexity scales quadratically with the number of image patches, highlighting the need for more efficient alternatives. To address this challenge, we propose $\FMM$, a novel framework that integrates Selective State-Space Models (SSMs), transformer-based attention, and expert-driven feature refinement into a unified architecture. $\FMM$ extends the MambaVision backbone by introducing the Sequential Mixture of Experts (SeqMoE) mechanism through its customized $\SMB$ block. The $\SMB$ is a modified MambaVision block that enhances representation learning in high-resolution mammographic images by enabling content-adaptive feature refinement. These blocks are integrated into the deeper stages of MambaVision, allowing the model to progressively adjust feature emphasis through dynamic expert gating, effectively mitigating the limitations of traditional Transformer models. Evaluated on the CBIS-DDSM benchmark dataset, $\FMM$ achieves superior classification performance across all key metrics while maintaining computational efficiency. 

\end{abstract}

\section{Introduction}
\vspace{-.05in}
Breast Cancer (BC) is the most commonly diagnosed cancer among women worldwide~\cite{sung2021global}. Regular screening, early diagnosis, and improved treatment planning~\cite{ghassemi2024novel, rooks2023po91} are key factors in reducing mortality rates and associated disease burden~\cite{mcintire2023novel}. This underscores the critical importance of advancing accurate and reliable diagnostic solutions to enhance early detection and improve patient survival.
Mammography is the primary screening method for BC and typically relies on analyzing images from multiple views, i.e., CranioCaudal (CC) and MedioLateral Oblique (MLO). Multi-view mammography facilitates the identification of subtle abnormalities that may not be visible in a single view. The complexity of these images, however, makes manual interpretation particularly challenging due to subtle tissue differences, overlapping anatomical structures, and variations in breast density~\cite{pisano2005diagnostic}, all of which can obscure malignancies and contribute to missed detections and false positives. As a result, there has been a growing interest in the development of Computer-Aided Diagnosis (CAD) systems~\cite{katzen2018review} to automate mammogram analysis and improve diagnostic accuracy.

\vspace{.025in}
\noindent
\textbf{\textit{Literature Review:}} In recent years, Artificial Intelligence (AI)-powered diagnosis and prognosis models have emerged as promising complementary tools to support radiologists in advanced mammographic analysis. Early Deep Learning (DL) models for BC diagnosis and prognosis~\cite{sun2019multi,khan2019multi, shen2019deep, baccouche2022integrated} were primarily based on Convolutional Neural Networks (CNNs), which are well-suited for local feature extraction. However, CNNs face limitations in modeling long-range dependencies across high-resolution images, restricting their ability to capture global contextual relationships critical for accurate and reliable diagnosis. Consequently, there has been a paradigm shift toward Transformer-based architectures~\cite{sarker2024mv} for multi-view mammography, addressing these limitations by modeling long-range dependencies through self-attention mechanisms. 
Building on this, the Multi-Scale Multi-View Swin Transformer (MSMV-Swin)\cite{bayatmakou2025} demonstrated the value of combining fine-grained tumor features with broader anatomical context, achieving robust classification performance.
Furthermore,
hybrid models\cite{boudouh2024advancing, allaoui2024hybridmammonet} 
have emerged, integrating the strengths of CNNs and Transformers for more efficient multi-view feature fusion. However, the computational complexity of Transformer architectures scales quadratically with the number of image patches, making them inefficient for large mammographic datasets.

This paper aims to address the aforementioned gap by capitalizing on recent advances in sequence modelling and Selective State-Space Models (SSMs), particularly the Mamba framework~\cite{gu2023mamba}. SSMs offer linear-time complexity, making them a scalable solution for processing long sequences such as mammographic image patches. Several Mamba variants such as Vision Mamba~\cite{Zhu2024vision}, Vmamba~\cite{liu2024vmamba}, Multi-scale vMamba~\cite{shi2024multi}, MedMamba~\cite{yue2024medmamba}, and most recently Dynamic Vision Mamba~\cite{wu2025dynamic} have been developed for visual tasks. To effectively balance local and global feature extraction, Mamba variants typically combine SSMs with convolutional layers. MambaVision~\cite{hatamizadeh2025mambavision}, a recently introduced architecture, further enhances this balance by integrating SSMs with transformer blocks, creating a hybrid model more suitable for images such as mammograms. Its linear scaling properties and robust multi-scale feature extraction capabilities make MambaVision a strong candidate backbone for mammography, where capturing both fine-grained tumor details and broader anatomical context is crucial. To the best of our knowledge, this work is among the first to adapt MambaVision for multi-view mammogram classification. 

\vspace{.025in}
\noindent
\textbf{\textit{Contributions:}}
Building on MambaVision, we propose $\FMM$, a framework that integrates a Sequential Mixture of Experts (SeqMoE) mechanism into the deeper stages of MambaVision for adaptive feature refinement. 
To adapt the MambaVision model for multi-view mammogram classification, the following strategies can be pursued: \textit{(i) Linear Probing}, where the model's pre-trained weights are frozen and embeddings from the last layer are passed to linear layer(s) for prediction; \textit{(ii) Full Fine-Tuning}, where the entire MambaVision model is initialized with pre-trained weights and then updated using the mammography dataset; and \textit{(iii) Partial Tuning}, where only selected internal layers (e.g., via low-rank adapters~\cite{hu2022lora}) are unfrozen and fine-tuned. We propose to go one step beyond (iii): in addition to fine-tuning the patch embedding layer, we modify the internal Mamba blocks of the MambaVision architecture. Specifically, we introduce the $\SMB$ block, which is constructed based on the proposed SeqMoE mechanism. In summary, the paper makes the following contributions.
\begin{itemize}
\item We propose the $\FMM$, a dual-stream framework for multi-view mammogram classification that integrates the proposed SeqMoE mechanism into MambaVision. The $\FMM$ introduces a novel architectural design that combines SSM, transformer-based attention, and expert-driven feature refinement into a unified framework. 
\item We introduce the $\SMB$ block, a modified MambaVision block built upon SeqMoE to enhance representation learning in high-resolution mammographic images. The SeqMoE mechanism enables the model to dynamically adjust feature emphasis depending on the content, enhancing robustness and adaptability in diverse clinical settings.
\end{itemize}
Evaluated on the CBIS-DDSM dataset, $\FMM$ achieves superior classification accuracy compared to its state-of-the-art counterparts while maintaining computational efficiency. 
The remainder of the paper is organized as follows. Section~\ref{sec:back} presents the background information. Section~\ref{sec: methodology} details the $\FMM$ framework together with its $\SMB$ block and SeqMoE gating mechanism. The experimental setup and results are included in Section~\ref{sec: experiments}. Finally, Section~\ref{sec: conclusion} concludes the paper.

\section{Background}\label{sec:back}
In this section, the utilized dataset and the MambaVision Backbone are briefly presented.


\subsection{Dataset}
To evaluate the proposed $\FMM$, the Curated Breast Imaging Subset of the Digital Database for Screening Mammography (CBIS-DDSM) dataset~\cite{lee2017curated} is used.
It is a widely recognized benchmark for mammography research, including $10,239$ images, comprising whole mammograms, cropped Regions of Interest (ROIs), and masks for masses and calcifications across diverse patient cases.
We used the train-test split recommendation for the dataset, excluding missing views and considering only one scan per breast. We selected $653$ mammography examinations focusing solely on masses, each representing a single breast. These examinations were split into $504$ training and $149$ testing cases. Each examination consisted of four images: cropped and whole mammograms from both CC and MLO views.
We uniformly flipped the images to address variability between left and right breast scans.
The $SSM$ framework uses cropped images to focus on tumor-specific regions and whole mammograms to capture comprehensive breast anatomy, potentially enhancing multi-view classification performance.

\subsection{MambaVision Backbone}\label{sec:mamba_vision}
The SSM-based models, including Mamba, are built upon classical state-space models from signal processing and control theory. MambaVision~\cite{hatamizadeh2025mambavision} is a hierarchical vision model that integrates convolutional layers, SSMs, and transformer-based self-attention blocks to efficiently extract local and global features. Unlike traditional ViTs with quadratic complexity, MambaVision achieves improved scalability by using SSMs to model long-range dependencies with linear-time complexity. This provides scalability due to the use of linear-time processing. In addition, compared to models based on self-attention, linear-time complexity of MambaVision ensures efficiency through a lower GPU memory usage~\cite{gu2023mamba}. Furthermore, by using SSMs, it can provide a global context by capturing long-distance features.
In brief, MambaVision processes an input image hierarchically in the following steps: (i) \textit{Patch Embedding}: Convolutional layers transform the input image into an initial feature representation; (ii) \textit{Stages 1 and 2}: Residual convolutional blocks extract low-level spatial features at high resolutions; (iii) \textit{Stages 3 and 4}: A hybrid architecture combines MambaVision Mixer blocks, which utilize SSMs for long-range modeling, with self-attention blocks to refine local and global features. Such a hierarchical design enables efficient multi-scale feature extraction, with convolutional blocks in early stages and a combination of MambaVision Mixer and self-attention blocks in deeper stages.

\section{The $\FMM$ Framework} \label{sec: methodology}
\setlength{\textfloatsep}{0pt}
\begin{figure*}
    \centering
    \includegraphics[width=.7\textwidth]{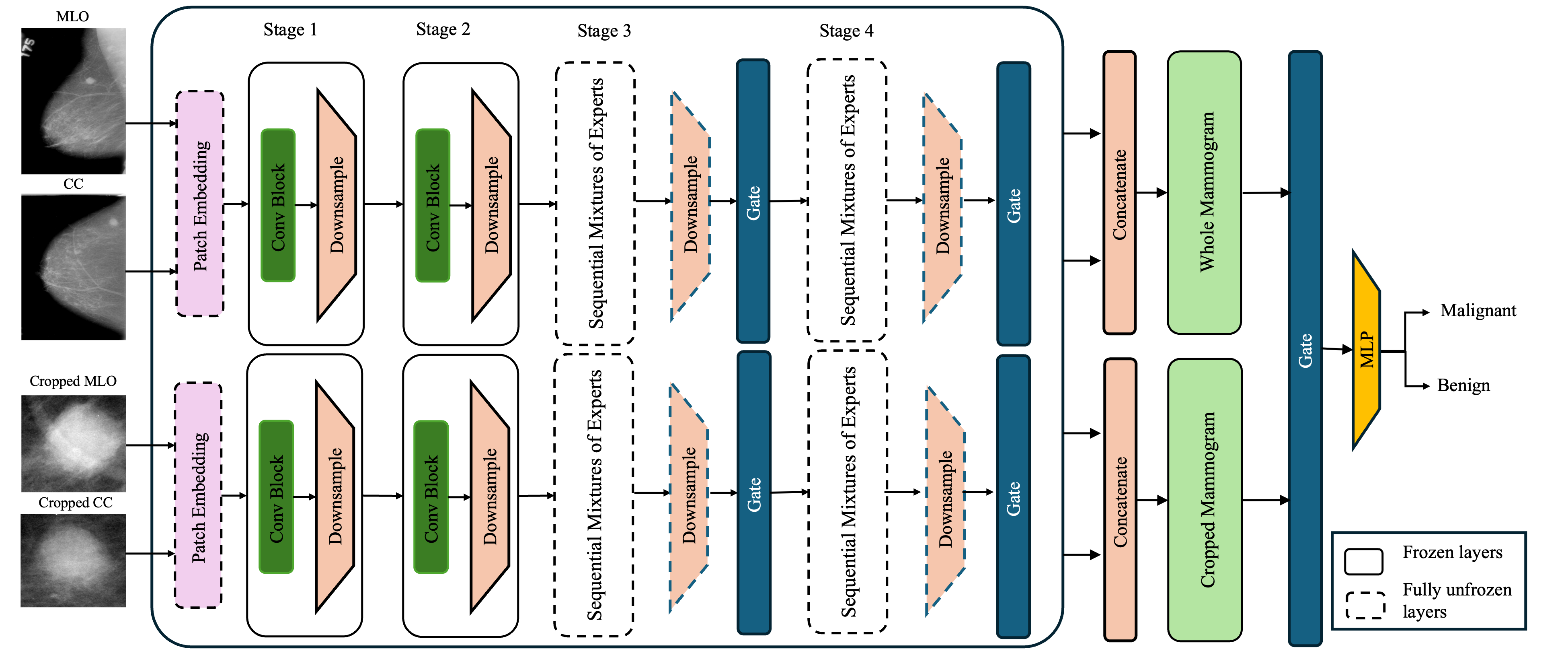}
    \vspace{-.1in}
    \caption{Architecture of the Proposed Model}   \label{fig:modified_mambaVision_architecture}
     \vspace{-.2in}
\end{figure*}
\begin{figure}
    \centering
    \includegraphics[width=\columnwidth]{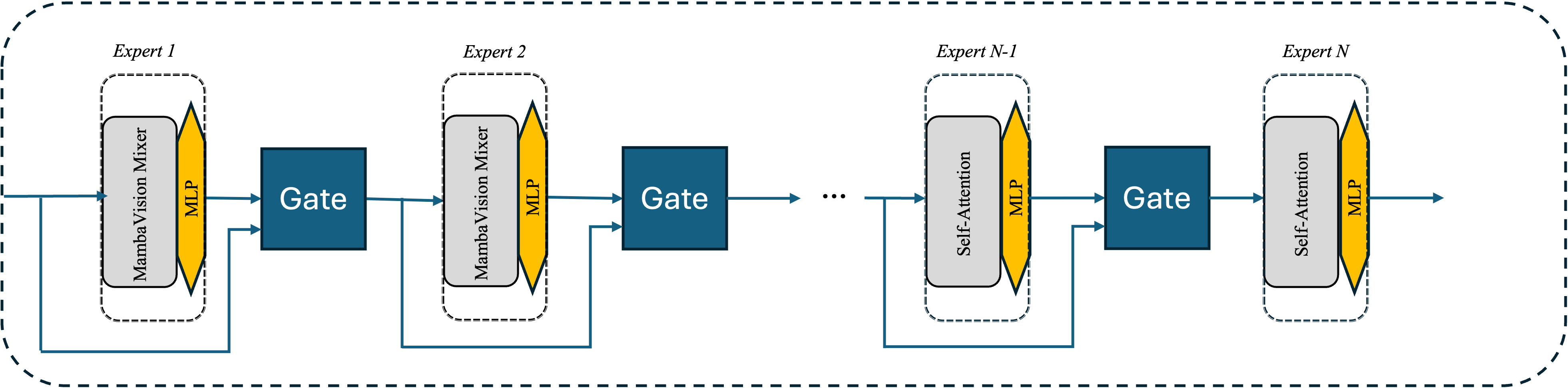}
    \vspace{-.25in}
    \caption{Micro-Architecture of SeqMoE: MambaVision Mixer and self-attention blocks with iterative gating after each expert.}
    \label{fig: Seq_architecture}
\end{figure}
In this section, we present the proposed $\FMM$ architecture that adopts the MambaVision backbone for the task of multi-view mammogram classification. The overall architecture of the $\FMM$ framework is shown in Fig.~\ref{fig:modified_mambaVision_architecture}, with its detailed micro-architecture depicted in Fig.~\ref{fig: Seq_architecture}.

\subsection{Patch Embedding}\label{sec:Patch_Embedding}
Let the input mammogram dataset be represented by $\mX\i \in\mathbb{R}^{H\times W\times C}$, for ($1 \leq  i \leq N_{\text{I}}$),  where $N_{\text{I}}$ is the number of input mammogram images. $H$ and $W$ represent the mammogram's height and width, and $C$ denotes the number of channels (typically, $C=1$ for grayscale mammograms).

To process an image in the $\FMM$ pipeline, it is first divided into non-overlapping patches and flattened. More specifically, $\mX\i$, for ($1 \leq  i \leq N_{\text{I}}$), is divided  into $N_{\text{P}}$ number of patches denoted by $\Z\ik$, for ($1 \leq k \leq N_{\text{P}}$), with each of size ($P \times P$). This results in $N_{\text{P}} = H.W/P^2$ total number of patches. Each patch $\Z\ik \in \mathbb{R}^{P \times P}$ is flattened into the patch-embedding vector denoted by $\z\ik \in \mathbb{R}^{P^2}$. 
A shared linear projection is then applied as follows
\begin{eqnarray}
\p\ik = \W_{\text{P}} \z\ik+\b_{\text{P}},\label{eq:PEtoken}
\end{eqnarray}
where $\mathbf{W}_{\text{P}} \in \mathbb{R}^{d_{\text{PE}} \times P^2}$ is the shared linear projection matrix, and $d_{\text{PE}}$ is the dimension of Patch Embedding (PE) vector $\p\ik$. Stacking patch embeddings results in the PE sequence. 
\begin{eqnarray}
\P\i = [\p_{i,1}, \ldots \p_{i,k}, \dots, \p_{i, N_{\text{P}}}] \in \mathbb{R}^{N_{\text{P}} \times d_{\text{PE}} }.\label{eq:U}
\end{eqnarray}
This completes construction of the patch embedding input.

\vspace{-.01in}
\subsection{$\SMB$ Blocks}\label{sec:SMB}
%
Let $N_{\text{L}}$ denote the number of $\SMB$ blocks $\mathcal{B}\ul_{\text{SecM}}$, stacked in series, for ($1 \leq \ell \leq N_{\text{L}}$). The input tokens of each $\SMB$ block $\mathcal{B}\ul_{\text{SecM}}(\cdot)$ corresponding to input mammogram $\mX\i$ is denoted by $\T\ul\i \in \mathbb{R}^{N_{\text{P}} \times d_{\text{PE}}}$. At the first stage of each $\SMB$ block, a shared linear projection is applied to its input tokens to construct Input Embedding (IE) of $d_{\text{IE}}$. In other words, the input token sequence $\T\ul\i$ is projected to a new latent space $\X\ul\i$ of dimension $d_{\text{IE}}$, i.e.,
\begin{equation}
\X\i\ul = \W_{\text{in}} \T\i\ul + \b_{\text{in}}, \quad \X\i\ul \in \mathbb{R}^{N_{\text{P}} \times d_{\text{IE}}}\label{eq:IE}
\end{equation}
Each of the $N_{\text{P}}$ embedding tokens $\x\k\ul \in \mathbb{R}^{d_{\text{IE}}}$, ($1\leq k \leq N_{\text{P}}$) corresponds to one image patch of $\mX\i$. Similar to Eq.~\eqref{eq:U}, $\X\i$ can be written as $[\x_{i,1}, \ldots \x_{i,k}, \dots, \x_{i, N_{\text{P}}}]^T \in \mathbb{R}^{N_{\text{P}} \times d_{\text{IE}}}$.
%
%
For simplicity of the notation, we drop the superscript $\ul$, which will be followed below as well.  The $\SMB$ block is formulated as an SSM and is applied to its input token sequence $\X\i$. Without loss of generality, we consider the following general discrete-time linear SSM
\begin{eqnarray}
\!\!\!\!\!\!\!\!\!\!\!\!&&\!\!\!\!\!\!\!\! \text{State Model:} \qquad\qquad~~ \h\k = \A \h\pk + \B \x\pk \label{eq:Mstat}  \\
\!\!\!\!\!\!\!\!\!\!\!\!&&\!\!\!\!\!\!\!\! \text{Observation Model:} \qquad  \y\k = \C \h\k, \label{eq:Mobs}
\end{eqnarray}
where $\x\k \in \mathbb{R}^{d_{\text{IE}}}$ is the input token associated with patch (time index) $k$, for ($1\leq k \leq N_{\text{P}}$). Vector $\h\k \in \mathbb{R}^{d_{\text{HS}}}$ is the hidden state of dimension $d_{\text{HS}}$, and $\y\k \in \mathbb{R}^{d_{\text{IE}}}$ is the output token. Matrices $\A, \B, and \C$ are learnable (structured) state and observation model matrices of appropriate dimensions.

To process sequences efficiently, Mamba reformulates the SSM (Eqs.~\eqref{eq:Mstat}, and~\eqref{eq:Mobs}) into a convolutional operation. More specifically, for an input sequence $\X\i$ with size of $N_{\text{P}}$, a global convolution kernel $\K$ is used for computing the output of Eq.~\eqref{eq:Mobs} as follows
\begin{eqnarray}
\Y\i = [\y_{i,1} \ldots \y\ik, \ldots,\y_{N_{\text{P}}}]^T = \K^T * \X\i  \label{eq:convY}
\end{eqnarray}
where $\Y\i, \K \in \mathbb{R}^{N_{\text{P}} \times d_{\text{IE}}}$, and $*$ denotes 1D convolution along the sequence (temporal) axis. The convolutional kernel $\K$ in Eq.~\eqref{eq:convY} is expressed as follows over time steps ($0 \leq k \leq N_{\text{P}}$) 
\begin{equation}
\K = \C\A^{k}\B =  \left[ \C \B, \, \C \A \B, \, \, \dots, \, \C \A^{N_{\text{P}}-1} \B \right].
\end{equation}
%
%
Kernel $\K$ is a parameterized/derived convolution kernel computed from discretized state-space parameters ($\A, \B, \C$). Please note that in Eq.~\eqref{eq:convY} the convolution length is equal to $N_{\text{P}}$, which is the number of input patches. 
The kernel $\K$ can be computed in closed form or approximated efficiently using structured parameterizations, such as low-rank factorizations or diagonal state matrices. 

To construct the output $\Y\i^{\text{SM}}$ of the $\SMB$ block, first, selective modulation is applied on Eq.~\eqref{eq:convY} as follows
\begin{eqnarray}
\Y\i^{\text{MD}} &=& \G \odot \Y\i, \\
\G &=& \sigma(\W_g \X\i + \b_g),
\end{eqnarray}
where the selective filtering gate $\G \in \mathbb{R}^{N_{\text{P}} \times d_{\text{IE}}}$ modulates the SSM output $\Y\i$ element-wise through sigmoid activation $\sigma(\cdot)$. The modulated representation $\Y\i^{\text{MD}}$ controls what information is retained. Such a modulation allows the model to learn which time/position step to enhance or suppress, analogous to attention. The final output of the $\SMB$ block is projected back to the original embedding space as 
\begin{equation}
\Y\i^{\text{SM}} = \W_{\text{out}} \Y\i^{\text{MD}} + \b_{\text{out}}, \quad \Y\i^{\text{SM}} \in \mathbb{R}^{N_{\text{P}} \times d_{\text{PE}}}, \label{eq:Mout}
\end{equation}
which maps the latent representation back to the original embedding space. The output of a Mamba block for IE $\X\i$ (which is computed from the block's input $\T\i\ul$ via Eq.~\eqref{eq:IE}) is denoted by $\mathcal{B}_{\text{M}}(\X\i) \!=\! \text{LayerNorm} ( \X\i \!+\! \Y\i^{\text{SM}})$. Expanding $\mathcal{B}_{\text{M}}(\X\i)$ based on Eqs.~\eqref{eq:PEtoken}-\eqref{eq:Mout} results in 
\begin{eqnarray}
&& \!\!\!\! \!\!\!\!\!\!\!\!\! \mathcal{B}_{\text{M}}(\X\i\ul)  \!=\! \text{LayerNorm} \Big( \X\i\ul + \Big. \nonumber\\
&&\!\!\!\! \!\!\!\! \W_{\text{out}} \big[ \left( \sigma(\W_g \X\i\ul + \b_g) \right) \odot \K *  \X\i\ul \big] + \b_{\text{out}}\Big)  
\end{eqnarray}


\vspace{.1in}
\noindent
\textbf{\textit{The Sequential Mixture of Experts (SeqMoE):}} The SeqMoE is a custom gating mechanism $\mathcal{G}(\cdot)$ placed after each block, for ($1 \leq \ell \leq N_{\text{L}}$), where as stated previously, $N_{\text{L}}$ is the number of blocks (layers) stacked in series. The gating mechanism $\mathcal{G}(\cdot)$ is a binary gating with two inputs: (i) The $\ell^{\text{th}}$ block's input, which is the output of the previous block $\mathcal{B}_{\text{SecM}}(\X^{(\ell-1)}_i)$, and; (ii) Its intermediate output $\mathcal{B}_{\text{M}}(\X\ul_i)$. The gating mechanism $\mathcal{G}(\cdot)$ is defined as follows
\begin{eqnarray}\label{eq:Gating}
\lefteqn{
\mathcal{G}\left(
\mathcal{B}_{\text{M}}(\X\ul_i), 
\mathcal{B}_{\text{SecM}}(\X^{(\ell-1)}_i)
\right) = 
\text{Softmax} \Big( 
\W_{\text{out}}^{\mathcal{G}} } \\
&& \cdot \text{ReLU} \left( \W_{\text{in}}^{\mathcal{G}} 
\left[
\mu\left(\mathcal{B}_{\text{M}}(\X\ul_i)\right), \,
\mu\left(\mathcal{B}_{\text{SecM}}(\X^{(\ell-1)}_i)\right)
\right] 
\right) 
\Big), \nonumber
\end{eqnarray}
where the output of gate $\mathcal{G}$ is a scalar weight representing the importance of the latest expert block output. Function $\mu(\cdot)$  provides mean pooled representation of its input, and \(\W_{\text{in}}^{\mathcal{G}}, \W_{\text{out}}^{\mathcal{G}}\) are learnable SeqMoE gating parameters. The notation \([\,\cdot\,,\,\cdot\,]\) denotes concatenation of the two inputs.
This formulation enables adaptive depth-wise fusion within each stage of the network. By reusing the same gating network across expert blocks in a given stage, the model maintains parameter efficiency. Meanwhile, using separate gating functions in different stages supports more flexible and stage-specific control over the sequential refinement of representations.
The final output of the $\ell^{\text{th}}$ block becomes a gated interpolation between its input and output, i.e.,
\begin{equation}
\mathcal{B}_{\text{SecM}}(\X\i\ul) = \mathcal{G}\, \mathcal{B}_{\text{M}}(\X\i\ul) + (1 - \mathcal{G})\, \mathcal{B}_{\text{SecM}}(\X\i^{(\ell-1)}).
\end{equation}
This is an adaptive residual connection controlled by $\mathcal{G}$, similar in spirit to a GRU-style gate, but without recurrence. The gating mechanism $\mathcal{G}$ learns how much transformation to apply at each position in the sequence.

%
It is worth noting that the introduced gating mechanism shares conceptual DNA with MoE. In the classical MoE,  multiple expert modules operate in parallel, and the gating function forms a weighted combination of experts' outputs. Instead of mixing multiple experts at once, in the SeqMoE mechanism, each block $\mathcal{B}\ul_{\text{SecM}}(\cdot)$ is treated as an expert paired with its previous expert, resulting in a pairwise expert mixture. One expert transforms the input once, while the other transforms the input twice (stacked depth). The gate decides at each spatial/temporal position whether a single or double transformation is more appropriate.
Introduction of dynamic depth routing allows the model to adaptively choose how deeply to process each input based on its content. Rather than relying on a fixed architecture, the model can decide in real-time whether to apply a single transformation or a deeper sequence of operations, similar to dynamic unrolling or early exiting, but inherently built into the network structure. In other words, the SeqMoE architecture is an MoE in the depth dimension, where the gating mechanism selects between different levels of computation.

\subsection{$\FMM$ Architecture}
\begin{table*}[t!]
    \centering
    \renewcommand{\arraystretch}{1.1} 
    \setlength{\tabcolsep}{6pt} 
    \caption{Performance Metrics Across Mamba-based Models}
     \vspace{-.1in}
    \label{tab: Mamba-comparisons}
    \begin{tabular}{lccccc}
        \toprule
        \textbf{Model Type} & \textbf{Accuracy} & \textbf{AUC} & \textbf{F1 Score} & \textbf{Sensitivity} & \textbf{Specificity} \\
        \midrule
        MedMamba (Ultrasound) & 0.6620 & 0.6945 & 0.5944 & 0.6045 & 0.7020\\
        MedMamba (CPNXray) & 0.6921 & 0.7106 & 0.5933 & 0.5480 & 0.7922\\
        Vmamba S & 0.8054 & 0.8510 & 0.7680 & 0.8136 & 0.8000 \\
        Vmamba T & 0.7919 & 0.8480 & 0.7597 & 0.8305	 & 0.7667 \\
        Vmamba B & 0.8121 & 0.8584 & 0.7667 & 0.7797 & 0.8333 \\
        Mamba Vision B & 0.7535 & 0.8340 & 0.7054 & 0.7203 & 0.7765 \\
        Mamba Vision T1 & 0.7987 & 0.8427 & 0.7500 & 0.7627 & 0.8222 \\
        MambaVision T2 & 0.7778 & 0.7257 & 0.8436 &  0.7175 &  0.8196 \\ 
        MambaVision L & \textbf{0.8255} & 0.8774 & 0.7969 & 0.8644 & 0.8000 \\
        \bottomrule
    \end{tabular}
    \vspace{-.1in}
\end{table*}
\begin{table*}[t!]
    \centering
    \renewcommand{\arraystretch}{1.1} 
    \setlength{\tabcolsep}{6pt} 
    \caption{Performance Metrics using Multiple Experts Strategy}
    \vspace{-.1in}
    \label{tab: experts}
    \begin{tabular}{lccccc}
        \toprule
        \textbf{Model Type} & \textbf{Accuracy} & \textbf{AUC} & \textbf{F1 Score} & \textbf{Sensitivity} & \textbf{Specificity} \\
        \midrule
 (a) MambaVision  with 2-Streams (Crop + Whole) & 0.8322 & 0.8791 & 0.8092 & 0.8983 & 0.7889 \\
 (b) MambaVision  with 4-Streams (Per View) & 0.8054 & 0.8740 & 0.7642 & 0.7966 & 0.8111 \\
(c) MambaVision Mixers with 2-Streams & 0.8244 & 0.8631 & 0.8296 & 0.8750 & 0.7761\\
(d) \textbf{$\FMM$} & \textbf{0.8792} & \textbf{0.9249} & \textbf{0.8525} & \textbf{0.8814} & \textbf{0.8778} \\

                \bottomrule
    \end{tabular}
    \vspace{-.1in}
\end{table*}

As stated previously, the $\FMM$ architecture (shown in Fig.~\ref{fig:modified_mambaVision_architecture}) is constructed upon the MambaVision backbone consisting of four stages. The patch embedding layer is initialized with MambaVision values and then fully fine-tuned. Stages $1$ \& $2$ of the $\FMM$ are frozen instances from MambaVision to preserve general low-level representations. Stages $3$ \& $4$ are modified to adapt to mammographic features. Particularly, in Stages $3$ \& $4$, $\SMB$ blocks and SeqMoE gating are integrated as~follows
\begin{itemize}
\item Stage 3: It consists of $10$ experts including $5$ $\SMB$ blocks followed by $5$ self-attention blocks. The SeqMoE mechanism iteratively merges the outputs of these experts. 
\item Stage 4: A convolutional downsampling module is applied first to reduce spatial resolution before Stage $4$. Then $5$ experts are stacked in series, including $3$ $\SMB$ blocks followed by $2$ self-attention blocks. SeqMoE gating is again applied iteratively to mix the experts.
\end{itemize}
%

\vspace{.1in}
\noindent
\textbf{\textit{Dual-Stream Feature Fusion:}} The $\FMM$ architecture is a dual-stream fusion consisting of: (i)  \textit{Crop Stream,} that extracts fine-grained features from lesion-focused regions, and (ii) \textit{Mammogram Stream,} captures global anatomical structures from whole mammogram images.
Each stream is processed independently using a dedicated copy of the $\FMM$ architecture stages, which include $\SMB$ blocks and self-attention blocks generating feature embeddings for CC and MLO views. These are then concatenated into two representations, one associated with the Crop Stream and one for the Mammogram Stream. These embeddings are fused using a gating mechanism, resulting in the final feature vector, which is passed through a Multi-Layer Perceptron (MLP) for classification.

\begin{table*}[t!]
    \centering
    \renewcommand{\arraystretch}{1.1} 
    \setlength{\tabcolsep}{6pt} 
    \caption{\footnotesize Comparison with state-of-the-art. }
    \vspace{-.1in}
    \label{table:tsota}
    \begin{tabular}{lccccc}
        \toprule
        \textbf{Model} & \textbf{Accuracy} & \textbf{AUC} & \textbf{F1 Score} & \textbf{Sensitivity} & \textbf{Specificity} \\
        \midrule

        Score-based Generative Modeling~\cite{sarker2025can} & 0.6365 & 0.7175 & - & 0.5821 & 0.7283 \\
        MV-Swin-T~\cite{sarker2024mv} & 0.6863 & 0.7137 & - & - & - \\
        Multi-Patch Size Image Classifier~\cite{quintana2023exploiting} & 0.730 $\pm$ 0.019 & 0.809 $\pm$ 0.005 & - & - & 0.710 $\pm$ 0.020 \\
        DCTr (ResNet-18)~\cite{dos2024deep} & 0.727 $\pm$ 0.191 & - & - & - & - \\
        Mammo-Clustering~\cite{yang2024mammo} & 0.709 & 0.805 $\pm$ 0.020& 0.709 & - & - \\
HybridMammonet~\cite{allaoui2024hybridmammonet} & - & 0.8000 & 0.6500 & - & - \\
        ResNet-50~\cite{liao2024open} & 0.6824 & 0.7421 & 0.6538 & 0.6296 & - \\
         XFMamba~\cite{zheng2025xfmamba} &- & 0.76 $\pm$ 0.003 & -& - & - \\
         
        MSMV-Swin~\cite{bayatmakou2025} & 0.8032 & 0.8227 & 0.7507 & 0.7232 & 0.8588 \\  
        \textbf{$\FMM$} & \textbf{0.8696 $\pm$ 0.0109} & \textbf{0.9089 $\pm$ 0.0135} & \textbf{0.8396 $\pm$ 0.0124} & \textbf{0.8620 $\pm$ 0.0248} & \textbf{0.8746 $\pm$ 0.0220} \\
    \bottomrule
    \end{tabular}
    \vspace{-.1in}
\end{table*}
\begin{table*}[t!]
    \centering
    \renewcommand{\arraystretch}{1.1}
    \setlength{\tabcolsep}{6pt}
    \caption{Ablation Study on Key Components}
    \vspace{-.1in}
    \label{tab: ablation}
    \begin{tabular}{lcc}
        \toprule
        \textbf{Configuration} & \textbf{Accuracy} & \textbf{AUC} \\
        \midrule
        
        MambaVision Backbone (Whole Mammograms) & 0.7383 & 0.7748 \\
        MambaVision Backbone (Cropped Mammograms) & 0.7517 & 0.7960 \\
        MambaVision Backbone + Dual-stream Experts & 0.8322 & 0.8791 \\
        $\FMM$(SeqMoE + Dual-Stream Experts) & 0.8792 & 0.9249 \\
        
        \bottomrule
    \end{tabular}
    \vspace{-.2in}
\end{table*}

\section{Experimental Setup and Results}\label{sec: experiments}
In this section, we evaluate the performance of the proposed $\FMM$ framework based on the benchmark CBIS-DDSM dataset, comparing it with a range of Mamba-based baselines and state-of-the-art models. Accuracy, AUC, F1-score, sensitivity, and specificity are used as performance metrics.  
All models were trained using the AdamW optimizer with a learning rate between $5\times10^{-5}$ and $7\times10^{-5}$, and weight decay in the range of $1\times10^{-3}$ to $3\times10^{-3}$. Training was conducted for $100$ epochs with cosine annealing scheduling. Images were resized to ($224\times224$) pixels and augmented with random color jitter to improve robustness. A batch size of $8$ was used, with cross-entropy loss incorporating label smoothing and gradient clipping (max-norm $1.0$) to stabilize training. Dataset splitting was performed at the breast level to avoid patient-level data leakage.

\vspace{.05in}
\noindent
\textbf{\textit{Comparison with Mamba-based Models:}} Table~\ref{tab: Mamba-comparisons} report the performance of $\FMM$ alongside baseline models, including MedMamba~\cite{yue2024medmamba}, Vmamba variants~\cite{liu2024vmamba}, and multiple MambaVision configurations~\cite{hatamizadeh2025mambavision}. Feature fusion in the dual-stream architecture (crop+whole) was performed using concatenation and maximum pooling strategies. Among the implemented models, MambaVision-L yielded the best performance and is thus used as the selected backbone to construct $\FMM$.

\vspace{.05in}
\noindent
\textbf{\textit{Effect of Expert Configurations:}} To explore the impact of expert configurations, we evaluated several setups within the MambaVision-L architecture. Table~\ref{tab: experts} explores the following different strategies:
\begin{itemize} 
\item[(a)] A 2-stream configuration with MambaVision backbones per stream, with one stream for cropped images and another for whole mammograms.
\item[(b)] A 4-stream configuration of MambaVision backbones, with one stream per view (cropped CC, cropped MLO, whole CC, whole MLO). 
\item[(c)] A two-stream configuration of a modified MambaVision where attention blocks are excluded, $5$ $\SMB$ experts in Stage $3$, and $3$ $\SMB$ experts in Stage~$4$.
\item[(d)] The $\FMM$  with $10$ experts in Stage $3$ ($5$ $\SMB$ blocks and $5$ self-attention blocks), and $5$ experts in Stage $4$ ($3$ $\SMB$ blocks, and $2$ self-attention blocks), combined with a dual-stream setup. 
\end{itemize}
The $\FMM$ model, which combines SeqMoE and the dual-stream design, outperformed all other configurations on all metrics, demonstrating the advantage of progressive expert-guided refinement.

\vspace{.05in}
\noindent
\textbf{\textit{Comparison with the State-of-the-Art:}} We compare $\FMM$ with recent state-of-the-art mammography models in Table~\ref{table:tsota}, including transformer-based methods, CNN hybrids and generative approaches. $\FMM$ achieved the highest scores in all reported metrics, highlighting its ability to effectively balance the local lesion details with global anatomical context.
The average performance metrics of the proposed model over $7$ runs are summarized in Table~\ref{table:tsota}. These results
are in the form of Mean Value $\pm$ standard deviation. Under normal assumptions with $95$\% confidence, our results are $0.8696 \pm 0.008$ for accuracy, $0.9089 \pm 0.01$ for AUC, $0.8396 \pm 0.009$ for F1 score,  $0.8620 \pm 0.0184$ for sensitivity, and specificity of $0.8746 \pm 0.0163$. 

\vspace{.05in}
\noindent
\textbf{\textit{Ablation Study:}}An ablation study was conducted to assess the impact of key architectural components in the $\FMM$ framework, with results presented in Table~\ref{tab: ablation}. The study evaluated the MambaVision backbone in a stand-alone setting using whole or cropped images. The configuration with whole mammograms yielded an accuracy of $0.7383$ and AUC of $0.7748$, while the cropped-image setup slightly improved these metrics to $0.7517$ and  $0.7960$, indicating the value of localized features. Additional experiments incorporated the dual-stream configuration, which improved the performance by effectively fusing local and global feature representations. The full $\FMM$ model, including the dual-stream architecture and the SeqMoE module, achieved the highest performance, with the best accuracy of $0.8792$ and AUC of $0.9249$. Removing the SeqMoE component resulted in a noticeable decline, reverting to the dual-stream-only metrics of $0.8322$ accuracy and $0.8791$ AUC. These findings validate the SeqMoE module's role in adaptive feature refinement, complementing the dual-stream design's ability to capture comprehensive image information. 
%
\section{Conclusion}\label{sec: conclusion}
In this paper, we proposed $\FMM$, an expert-guided multi-view mammogram classification framework that integrates the efficiency of MambaVision with the adaptability of Sequential Mixture of Experts (SeqMoE). The $\FMM$  effectively captures localized tumor details and a broader anatomical context. This is achieved by embedding SeqMoE layers in the deeper stages of MambaVision via the customized $\SMB$ block, and employing a dual-stream processing approach. The sequential expert selection mechanism ensures progressive feature refinement, dynamically adapting to mammographic variations. Extensive evaluations of the CBIS-DDSM dataset demonstrate that the $\FMM$ outperforms baseline Mamba-based models. A fruitful direction for future research is to investigate the generalization of the $\FMM$ framework across challenging scenarios such as missing views, variable breast density, limited tumor annotations, and multi-task learning for joint calcification and mass classification. 

\bibliographystyle{IEEEtran}

\end{document}